\documentstyle[preprint,tighten,aps,floats,epsfig]{revtex}

\def\t{\tau}
\def\e{\epsilon}

\newcommand{\be}{\begin{equation}}
\newcommand{\ee}{\end{equation}}
\newcommand{\bea}{\begin{eqnarray}}
\newcommand{\eea}{\end{eqnarray}}

\begin{document}

\draft

\title{Harmonic crossover exponents in O($n$) models with the pseudo-$\e$ expansion approach}

\author{
Pasquale Calabrese${}^{1}$ and  
Pietro Parruccini${}^2$}  
\address{$^1$ Rudolf Peierls Centre for Theoretical Physics, 1 Keble Road, 
Oxford OX1 3NP, United Kingdom.}
\address{$^{2}$ Dipartimento di Chimica Applicata e Scienza dei 
Materiali (DICASM), Universit\`a di Bologna, 
Via Saragozza 8, I-40136 Bologna, Italy.
}

\date{\today}

\maketitle

\begin{abstract}
We determine the crossover exponents associated with the 
traceless tensorial quadratic field, the third- and fourth-harmonic operators 
for O($n$) vector models by re-analyzing the existing six-loop fixed 
dimension series with the pseudo-$\e$ expansion. 
With this approach we obtain accurate theoretical estimates 
that are in optimum agreement with other theoretical and 
experimental results.

\end{abstract}


% ========================= BODY =========================

\section{Introduction}

For many years the critical behavior of O($n$) vector models has caught a lot 
of attentions since most of  the physical systems undergoing second-order phase 
transitions belong to the O($n$) universality classes (see Ref. \cite{rev-01}
 for a recent review). 
Thus a precise determination of universal quantities as critical exponents and
amplitude ratios has became necessary.
The critical behavior of these physical systems may be obtained by 
field theoretical investigations  based on the
Landau-Ginzburg-Wilson Hamiltonian
\begin{equation}
{\cal H}= \int d^dx\left[ \frac{1}{2}\partial_\mu \vec{\phi}\cdot \partial_\mu \vec{\phi}
+ \frac{1}{2} r \vec{\phi}\cdot \vec{\phi} + \frac{1}{4!}u
(\vec{\phi}\cdot \vec{\phi} )^2\right],
\label{H}
\end{equation}
where $\vec{\phi}(x)$ is a  $n$-component real field.
An interesting issue  is to determine how the critical properties are influenced by the addition of a perturbation term to
the Hamiltonian (\ref{H}) 
\begin{equation}
{\cal H}= \int d^dx\left[ \frac{1}{2}\partial_\mu \vec{\phi}\cdot \partial_\mu \vec{\phi}
+ \frac{1}{2} r \vec{\phi}\cdot \vec{\phi} + \frac{1}{4!}u
(\vec{\phi}\cdot \vec{\phi} )^2 + h_p {\cal P} \right],
\end{equation}
where $h_p(x)$ is an external field coupled to ${\cal P}(x)$.
In fact, if  ${\cal P}$ is an eigenoperator of the Renormalization Group (RG)
 transformations,  
the singular part of the Gibbs free energy becomes a scaling function in the 
limit of  reduced temperature $t\to 0$ and  $h_p \to 0$, and  can be written 
as
\begin{equation}
{\cal F}_{\rm sing}(t,h_p) \approx |t|^{d\nu}
 \widehat{\cal F} \left(h_p |t|^{-\phi_p}\right),
\end{equation}
where $\phi_p \equiv y_p \nu$ is the crossover exponent associated with the
perturbation ${\cal P}$ and $y_p$ is the RG 
dimension of ${\cal P}$.
Moreover, one usually defines the indices $\beta_p$ and 
$\gamma_p$ which describe the low-temperature singular behavior of the 
average $\langle {\cal P}(x)\rangle\sim |t|^{\beta_p}$ and of the 
suscettivity 
$\chi_{\cal P}= \int d^d x \langle {\cal P}(x){\cal P} (0) \rangle_c\sim t^{-\gamma_p}$. 
They satisfy the scaling relations
\begin{equation}
\beta_p = 2 - \alpha  - \phi_p, \qquad
\gamma_p = - 2 + \alpha + 2\phi_p.
\label{scalrel}
\end{equation}

Among the perturbation operators, particularly important from the 
experimental and phenomenological point of view are the so-called harmonic
ones \cite{Wegner-72,wz-75,rev} 
\bea
{\cal P}_{2}(x) &=& \phi_i(x) \phi_j(x) - \delta_{ij}\, \frac{1}{n} \,
\vec{\phi}(x)\cdot \vec{\phi}(x),\nonumber\\
{\cal P}_{3}(x) &=& \phi_a\phi_b\phi_c-\frac{\vec{\phi}\cdot \vec{\phi}}{n+2}(\phi_a\delta_{bc}
+\phi_b\delta_{ac}+\phi_c\delta_{ab}),\label{Tdef}\\
{\cal P}_{4}(x) &=&\phi_a\phi_b\phi_c\phi_d
-\frac{\vec{\phi}\cdot \vec{\phi}}{n+4}(\delta_{ab}\phi_c\phi_d+5 {\rm perm.})
+\frac{(\vec{\phi}\cdot \vec{\phi})^2}{(n+4)(n+2)}(\delta_{ab}\delta_{cd} +2 {\rm perm.})\,,\nonumber
\eea
called second, third, and fourth harmonic operator respectively.
In the following we will denote the crossover exponent of ${\cal P}_i$ as 
$\phi_i$ and its RG dimension as $y_i$.
Higher order harmonic operators are generally reputed to be irrelevant at the 
three-dimensional O($n$) fixed point \cite{wz-75}, thus we will
not consider them here.

The crossover exponent $\phi_2$ associated with the traceless tensor 
field ${\cal P}_{2}(x)$ reveals the instability of the O($n$)-symmetric
theory against anisotropy \cite{Wegner-72,FP-72,FN-74,Aharony-76}.
It characterizes the phase diagram at the multicritical point where 
two critical lines O$(n)$ and O$(m)$ symmetric meet. 
In some cases this gives rise to a critical theory with enlarged
O$(n+m)$ symmetry \cite{cpv-02,PJF-74,Fisher-75,KNF-76}. 
Multicritical behavior arises in several different contexts in physics:
in anisotropic antiferromagnets in a uniform magnetic 
field \cite{FN-74,KNF-76},
in high $T_c$ superconductors (see e.g. Ref. \cite{ae-01} and note that 
in the SO(5) theory of superconductivity~\cite{Zhang-97} the multicritical 
point is effectively O(5) symmetric), 
in colossal magnetoresistance materials \cite{dag},
in certain theories of strong interactions \cite{strong}, etc.
This list is far from being exhaustive, we only quote some examples.
For the $XY$ model ($n=2$) the traceless tensor field ${\cal P}_2(x)$ and 
its correlation function are connected with the second-harmonic order 
parameter in density-wave systems~\cite{NA-97,LB-82}, which characterizes 
some liquid crystals at the nematic-smectic-$A$ 
transition\cite{NA-97,LB-82,ABBL-86,Girault-etal-89,Brock-etal-89,Garland-etal-93,%
Wu-etal-94,Aharony-etal-95}. 
The structure factor of the secondary order parameter ${\cal P}_2$,
that within RG methods has been determined in- \cite{NA-97,cpv-02a} 
and out-of-equilibrium \cite{cg-04},
has been experimentally measured using X-ray scattering 
techniques~\cite{Wu-etal-94,Aharony-etal-95}.
Finally the RG dimension $y_2$ enters in the study of crossover 
effects in diluted Ising antiferromagnets with $n$-fold degenerate ground 
state \cite{Fernandez-88}, in  models with random anisotropy \cite{RAM},
at certain quantum phase transitions \cite{sm-02}, and 
in other more complicated situations \cite{other}.
Even this list is far from being exhaustive.

The third-harmonic crossover exponent determines the phase 
diagrams at the smectic-A hexatic-B point in liquid
crystals \cite{ABBL-86,Aharony-etal-95}, in materials exhibiting
structural normal-incommensurate phase transitions \cite{nipt1,nipt2,nipt3}, 
and at the trigonal-to-pseudotetragonal transition \cite{amb-77}.
For $n=0$, $\phi_3$ is related to the partition function exponent 
of nonuniform star polymers with three arms \cite{dpv-03}.
Finally, for $n\geq2$ it determines the stability of O($n$) fixed points 
against $n+1$-state Potts-like perturbations \cite{potts} as it happens 
in the presence of stress or particular magnetic fields \cite{amb-77,mfd-76}.

The fourth-harmonic exponent $\phi_4$ is mainly related to the stability
of the O($n$) fixed point against fourth-order anisotropy \cite{cpv-02}, 
as e.g. the cubic one \cite{Aharony-76}.
It is worth mentioning that for $n=1$, even if the operators ${\cal P}_i(x)$
can not be defined through Eqs. (\ref{Tdef}), all the $\phi_i$ have non 
trivial values.
This fact has an interpretation in terms of a gas of $n$-colors loops (see 
e.g. \cite{n-82}) in the limit $n\rightarrow1$.

The exponents $\phi_i$ and $y_i$ with $i=2,3,4$ have been analyzed in the 
past with different theoretical methods, in the framework of the 
$\epsilon$-expansion~\cite{w-72,yy-74,k-81,ks-95,Wegner-72,wz-75,ABBL-86,cpv-02,dpv-03},
from the analysis of high-temperature expansion~\cite{PJF-74}, 
by means of Monte Carlo simulations \cite{bfms-98,ch-98,Hu-01}, 
in the large $n$ approach~\cite{wz-75,largen,G-02}, 
and in the fixed-dimension perturbative 
expansion~\cite{ABBL-86,cpv-02a,dpv-03}.

The aim of this paper is to determine the crossover exponents $\phi_i$ and 
the RG dimensions $y_i$ by re-analyzing the three-dimensional 
six-loop perturbative series~\cite{cpv-02a,dpv-03} with the
pseudo-$\e$ expansion trick~\cite{Ni}, since
in many cases this method provided  accurate results in the
determination of critical quantities (see, 
e.g., Refs.\cite{GZ-80,GZ-98,hol,fhy-00,altri}).
The idea behind this trick is very simple: one has to multiply the linear
term of the  $\beta$ function by a parameter $\t$, find the fixed points~(i.e.
the  zeros of the $\beta$ function) as series in $\t$ and analyze the results
as in the $\e$ expansion.
The critical exponents are obtained as series
in $\t$ inserting the fixed-point expansion in the appropriate RG functions.
With this trick the cumulation of the errors coming from the non-exact
knowledge of the fixed point and from the uncertainty in the resummation of 
the exponents is avoided. The obtained pseudo-$\e$ series are believed to be asymptotic, so an appropriate resummation is usually needed to have reliable estimates.~\cite{GZ-80} However,  it often happens that up to six-loop order,
 the pseudo-$\e$ expansions do not yet show their asymptotic nature, rendering 
effective a non-resummed evaluation based on simple Pad\'e.

%Note that now, differently from $\e$ expansion, only the value
%at $\t=1$ makes sense, since the original series are obtained in fixed 
%dimension $d=3$.

The paper is organized as follows.
In Sec. \ref{secqua} we analyze the quadratic crossover exponents, 
in Sec. \ref{seccub} the cubic and quartic ones.
In Sec. \ref{concl} we report all the pseudo-$\e$ estimates for harmonic 
exponents and compare them with other theoretical and experimental ones.

%it can be related to the instability against
%cubic anisotropy though $y_4=-\omega_{\rm cub}$ \cite{comm,cpv-02}, 
%already considered in the pseudo-$\e$ expansion in Ref. \cite{fhy-00}.

\section{Quadratic crossover exponents}
\label{secqua}

The six-loop RG perturbative series in the three-dimensional approach for the 
second harmonic operators were computed in Ref. \cite{cpv-02a}, whereas 
the $\beta$ function (necessary to find the stable fixed point) 
is reported for general $n$ in Ref. \cite{as-95}. 
By using these series, one obtains the pseudo-$\e$
expansion of all the second-harmonic exponents.

\begin{table*}[b]
\begin{center}
\caption{\small Pad\'e table for $\phi_2^{n=2}$ in pseudo-$\e$ expansion. 
The two integer numbers $N$ and $M$ denote the corresponding $[N/M]$ Pad\'e.
The location of the positive real pole closest to the origin 
is reported in brackets.}
\begin{tabular}{l|ccccccc}
& $N=0$ & $N=1$ &$N=2$ &$N=3$ &$N=4$ &$N=5$&$N=6$ \\
\hline
$M=0$&1 &1.1 &  1.14696   &1.16385 &1.17390 &1.17616&1.18073\\
$M=1$& $\underline{1.11111}_{[10]}$ & $\underline{1.18855}_{[2.1]}$ &$\underline{1.17332}_{[2.8]}$&$\underline{1.18868}_{[1.7]}$&
$\underline{1.17683}_{[4.4]}$ & $\underline{1.17166}_{[0.5]}$ &\\
$M=2$&$\underline{1.15870}_{[4.0]}$  &$\underline{1.17369}_{[2.7]}$ &$\underline{1.17925}_{[2.3]}$& $\underline{1.17952}_{[2.3]}$&$\underline{1.18113}_{[2.1]}$&&\\
$M=3$&$\underline{1.17021}_{[3.2]}$  &$\underline{1.19549}_{[1.4]}$ &$\underline{1.17952}_{[2.3]}$&$\underline{1.17920}_{[0.1]}$&&&\\
$M=4$ & $\underline{1.17818}_{[2.7]}$  &$\underline{1.17773}_{[2.7]}$ &$\underline{1.18127}_{[2.03]}$&&&&\\
$M=5$ & $\underline{1.17771}_{[2.7]}$ & $\underline{1.17814}_{[2.7]}$&&&&&\\
$M=6$&$\underline{1.18271}_{[2.2]}$&&&&&&\\
\end{tabular}
\label{phin2}
\end{center}
\end{table*}

We first consider the crossover exponent $\phi_2$. As a typical example, 
the perturbative expression in the parameter $\tau$ for $n=2$ reads  
\be
\phi_2^{n=2}= 1 + \frac{\t}{10} + \frac{317\t^2}{6750} + 0.01688\t^3 + 0.01005\t^4 +
   0.00227\t^5 
%\nonumber \\&&
+ 0.00457\t^6 +O(\t^7)\,.
\ee
At least up to the presented number of loops the series does not behave as 
asymptotic with factorial growth of coefficients.
Although the series has not alternating signs, which is a key point to 
ensure some kind of convergence, one can try to apply a simple Pad\'e 
summation. 
The results for $n=2$ are displayed in Table \ref{phin2}. All the 
approximants possess poles on the real positive axis.
Some of them are close to $\tau=1$ and the estimate of 
$\phi_2$ on their basis should be considered unreliable. 
Anyway some of these approximants have poles ``far'' from
$\t=1$, where the series must be evaluated. Thus one may expect
the presence of such poles not to influence the approximant at $\t=1$.
Indeed all such Pad\'e results are very close since lower orders.
Hereafter we choose as final estimate the average of those six-loop order 
Pad\'e without poles in $0\leq\t\leq2$, and as error bar we take the maximum 
deviation of the final estimate from the four-, five- and six-loop  Pad\'e.
The five-loop estimates are analogously obtained, 
considering the maximum deviations up to three loops.
%as the average of 
%five-loop approximants and the error as maximum deviation from 
%three, four and five-loop Pad\'e.
Within this procedure we obtain $ \phi_2^{n=2}=1.178(15)$ at five-loop 
and $1.181(7)$ at six-loop. Although in good agreement with other
theoretical estimates, we do not retain safe such estimates, since the
presence of so many poles in the Pad\'e can cause systematic deviations
from the actual value.
This can be traced back to the fact that the series has not alternating signs. 
To improve the estimates, one can try
to resum the series by means of the
Pad\'e-Borel-Leroy (PBL) method (see e.g. \cite{GZ-80}) 
or more advanced ones, but  the monotonic character of the signs of 
the series makes the majority 
of the approximants to be defective. 
The resulting few good approximants do not allow a  safe determination of 
the quantities analyzed.
The same scenario is found for all other values of $n$.

\begin{table*}[t]
\begin{center}
\caption{\small Pad\'e table for $y_2^{n=2}$ in pseudo-$\e$ expansion. 
The two integer numbers $N$ and $M$ denote the corresponding $[N/M]$ Pad\'e.
The location of the positive real pole closest to the origin is reported in
brackets. The final estimate is $y_2^{n=2}=1.763(4)$.}
\begin{tabular}{l|ccccccc}
& $N=0$ & $N=1$ &$N=2$ &$N=3$ &$N=4$ &$N=5$&$N=6$ \\
\hline
$M=0$& 2&1.8&  1.76089   &   1.76705 &1.76215 &1.76707 &1.76041 \\
$M=1$& 1.81818 & $\underline{1.75138}_{[5.1]}$ &1.76621&1.76432&
$1.76461$ & 1.76424&\\
$M=2$&1.77061  &1.76756&$\underline{1.76376}_{[9.8]}$& 1.76459&1.76443&&\\
$M=3$&1.76774 &$\underline{1.77550}_{[0.6]}$ & 1.76471 &$\underline{1.76433}_{[13]}$&&&\\
$M=4$ & 1.76322  &1.76505 &$\underline{1.76405}_{[3.4]}$&&&&\\
$M=5$ & $ \underline{1.76630}_{[4.4]}$ &1.76444&&&&&\\
$M=6$&1.76162&&&&&&\\
\end{tabular}
\label{y2}
\end{center}
\end{table*}

To achieve a reliable estimate of $\phi_2$, one has
to consider series that have alternating signs. This can be done by
considering the RG dimension $y_2=\phi_2/ \nu$.
The pseudo-$\e$ expansion of $y_2$ for general $n$ is
\bea 
y_2&=&2-\frac{2}{(n+8)} \t+ \frac{2( -392 - 78n + 5n^2)}{27(8+n)^3} \tau^2
\nonumber\\&
+&\frac{362.271 + 65.4612\,n + 16.8274\,n^2 + 6.59007\,n^3 + 0.192062\,n^4}{(8+n)^5}\t^3\nonumber\\&
-&\case{19417.6 + 10881.8\,n + 2401.97\,n^2 - 29.3771\,n^3 - 83.6197\,n^4 - 5.54754\,n^5 - 0.0986411\,n^6}{(n+8)^7}\t^4\nonumber\\&
+&\case{1.66461\,{10}^6 + 981069.\,n + 215076.\,n^2 + 30474.1\,n^3 + 8065.54\,n^4 + 1734.33\,n^5 + 132.809\,n^6 + 4.22169\,n^7 + 0.0573847\,n^8}{(8+n)^9}\t^5\nonumber\\&
-&\left[\case{1.54538\,{10}^8 + 1.25687\,{10}^8 n + 4.50851\,{10}^7n^2 + 
9.02640\,{10}^6 n^3 + 7.98957\, 10^5 n^4 - 84675.5 n^5 - 31129.6 n^6 - 
2974.43n^7 
} {(8+n)^{11}}\right. \nonumber\\&&\left.\case{
- 133.244n^8 - 3.26010 n^9 - 0.0369419 n^{10}}{(8+n)^{11}}\right]\t^6\,,
\eea
that has alternating signs for $n\alt 6$. In fact, we get a small number of 
Pad\'e with poles on the real positive axis,
as one may appreciate from Table \ref{y2} where the results for $n=2$ 
are displayed. 
The goodness of the  Pad\'e persists 
increasing $n$ up to $n\simeq 6$ while for higher $n$ the results get worse.
All the final data are shown in Table \ref{risultati}, where we also report 
the result for $n=16$, that has to be taken with care since in this 
case the series is not alternating in signs.

We resum the perturbative series by the PBL method too. 
The number of defective approximants is very low and one may obtain 
a different estimate of the quantity $y_2$. 
We consider the four-, five-, and six-loop approximants. The final estimate
 is the mean value between  the maximum (M) and the minimum (m) value found. 
The error is $(M-m)/2$.  During this average procedure, we discard  the approximants
that seem to be pathological.
The results found with this method (displayed in Table \ref{risultati})
are very stable  and always compatible with the Pad\'e ones,
but they have  smaller errors. Only for $n=16$ we are not able to give a PBL
 final result since most of the approximants are defective.
This great stability within the PBL resummation makes us decide to report 
as final estimates (Table \ref{compare}) the simple Pad\'e ones, in order to
 avoid underestimation of the uncertainties.

Exploiting the scaling relations (\ref{scalrel}), we can apply the 
previous procedure to characterize the critical exponents $\beta_2$ and 
$\gamma_2$. Unfortunately these series have no alternating signs for all 
values of $n$, resulting in a bad determination of their actual value.
In Table \ref{risultati} we display $\phi_2$, $\beta_2$, and $\gamma_2$ 
as obtained by using scaling relations from $y_2$ and the most 
accurate estimates of standard critical exponents in the O($n$) universality
class \cite{best}.

\section{Third- and fourth-harmonic crossover exponents}
\label{seccub}

\begin{table}[b]
\begin{center}
\caption{\small Pad\'e table for $y_3^{n=0}$ in pseudo-$\e$ expansion. 
The location of the positive real pole closest to the
origin is reported in brackets.}
\begin{tabular}{l|ccccccc}
& $N=0$ & $N=1$ &$N=2$ &$N=3$ &$N=4$ &$N=5$&$N=6$ \\
\hline
$M=0$& 1.5& 0.75  &  0.71875  & 0.73111   &0.72040   & 0.73290& 0.71383  \\$M=1$& 1 & 0.71739$_{[24]}$  & 0.72761   &0.72537 &
 0.72617 &  0.72535 & \\
$M=2$& 0.84706  & 0.73152  & 0.72529$_{[47]}$ &0.72603 &0.72573&&\\
$M=3$ & 0.78599   & 0.71920$_{[3.5]}$ &0.72621
 & 0.72572$_{[60]}$ &&&\\
$M=4$ &0.75533  &  0.73227  &   0.72530$_{[6.5]}$
  &&&&\\
$M=5$&0.74240&0.68973$_{[1.3]}$  &&&&&\\
$M=6$&0.73215 &&&&&&\\
\end{tabular}
\label{taby3}
\end{center}
\end{table}

In this section we consider the critical exponents of the third- and 
fourth-harmonic operators. 

The six-loop three dimensional series relevant for ${\cal P}_3$ were 
calculated in Ref. \cite{dpv-03}. 
Even in this case only the direct estimate of $y_3$ gives rise to a
reliable result, since the series for $\beta_3$, $\gamma_3$, and $\phi_3$
have not alternating signs.
The pseudo-$\e$ expansion of $y_3$ for general $n$ is
\bea
y_3&=&\frac{3}{2}-\frac{6}{n+8} \t-\frac{144+44n -14n^2}{9(n+8)^3}\t^2\nonumber\\&
&+\frac{ 404.981 + 281.073\,n + 78.7256\,n^2 + 20.9419\,n^3 + 0.613212\,n^4}{(n+8)^5}\tau^3\nonumber\\&
+&\case{ -22461.1 - 17805.3\,n - 3307.73\,n^2 + 417.392\,n^3 + 244.052\,n^4 + 
   15.6783\,n^5 + 0.289476\,n^6}{(n+8)^7} \tau^4\nonumber\\&
+&\case{ 1.67818\,{10}^6 + 1.44153\,{10}^6\,n + 420045.\,n^2 + 92283.5\,n^3 + 
   24863.3\,n^4 + 4900.83\,n^5 + 366.716\,n^6 + 11.7112\,n^7 + 0.162327\,n^8}{(n+8)^9} \tau^5\nonumber\\&
-&\left[\case{1.63843\,{10}^8 + 1.78385\,{10}^8 n + 7.35887\,{10}^7 n^2 + 
1.53121\,{10}^7 n^3 + 968499. n^4 - 318541. n^5 - 87011.6 n^6 - 
   8034.60n^7 
}{(n+8)^{11}}\right.\nonumber\\&&\left. \case{
- 360.085 n^8- 8.93464 n^9 - 0.102502 n^{10}}{(n+8)^{11}}\right]\t^6\,,
\eea
which has alternating signs for $n\alt 5$.
To show the goodness of the Pad\'e summation we report in Table \ref{taby3}
the data for $n=0$.
The final estimate from this table is $y_3=0.725(29)$ 
[$y_3=0.731(35)$] at six-loop [five-loop]. These estimates are obtained using
 the procedure outlined in the previous section.
Similar good Pad\'e tables are found for higher values of $n$, up to 
$n\simeq 5$. 
All the final results are reported in Table \ref{risultati}, where  we also show the PBL ones for a comparison. Again the PBL uncertainty seems  too small to be considered safe.

It is worth noting that for the partition function exponent $p$ of 
non-uniform star polymers with three arms \cite{dpv-03}, 
we obtain the pseudo-$\e$ 
series
\be
p=3(\gamma+\nu)/2+\phi_3=
3  + \frac{\t^2}{16} - 0.009278\,\t^3 + 0.005889\,\t^4 - 0.014350\,\t^5 + 0.015458\,\t^6\,,
\ee 
which (by means of simple Pad\'e) leads to $ p=3.055(11)$. This value 
compares well with other estimates \cite{dpv-03} and with the one
obtained from $y_3$ and the most accurate theoretical estimates of $\gamma$ 
and $\nu$ \cite{best} leading to $p=3.043(18)$.

\begin{figure}[t]
\centerline{\epsfig{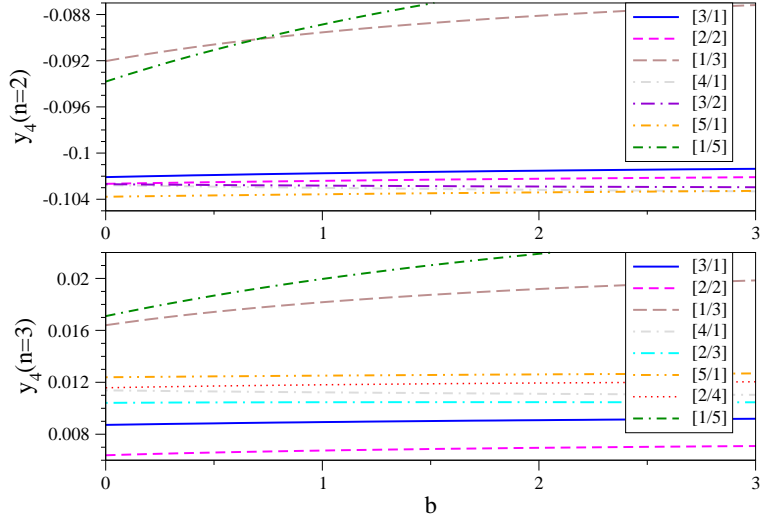}}
\vspace{5mm}
\caption{(Color on-line) Non-defective four-, five, and six-loop  PBL approximants 
for the RG dimension $y_4$ for $n=2,3$. }
\label{fig}
\end{figure}

Finally, let us consider the fourth-harmonic exponent.
It has been shown in Refs. \cite{comm,cpv-02} that the RG dimension
$y_4$ is related to the exponent characterizing the stability of the 
O($n$) fixed point against cubic anisotropy. 
Thus we can use the six-loop series of the cubic model \cite{cpv-00}
to obtain the pseudo-$\e$ expansion 
\bea
y_4&=&1-\frac{12}{n+8}\t+\frac{4( 680 + 62\,n + 23\,n^2)}{27(8+n)^3}\t^2
\nonumber\\&
+&\frac{-5200.56 - 777.127\,n + 33.2649\,n^2 + 35.7448\,n^3 + 1.25111\,n^4}{(n+8)^5}\t^3\nonumber\\&
+&\case{ 328835. + 90677.3\,n + 10871.8\,n^2 + 1595.82\,n^3 + 446.805\,n^4 + 
   28.9975\,n^5 + 0.574653\,n^6}{(n+8)^7}\t^4\nonumber\\&
+&\case{ -3.06135\,{10}^7 - 1.35901\,{10}^7\,n - 2.55849\,{10}^6\,n^2 - 
   179042.\,n^3 + 24766.4\,n^4 + 8290.96\,n^5 + 672.157\,n^6 + 22.2622\,n^7 + 
   0.318104\,n^8}{(n+8)^9}\t^5\nonumber\\&
+&\left[\case{3.18214\,{10}^9 + 1.83869\,{10}^9\,n + 4.52163\,{10}^8\,n^2 + 
   5.94633\,{10}^7\,n^3 + 5.79335\,{10}^6\,n^4 + 913345.\,n^5 + 
 161861.\,n^6 + 14567.2\,n^7
}{(n+8)^{11}}\right.\nonumber\\&& \left.\case{
+ 669.512\,n^8 + 17.0980\,n^9 + 0.199456\,n^{10}}{(n+8)^{11}}
\right] \t^6.
\label{y4eq}
\eea
From a general analysis \cite{rev-01,cpv-02}, it is known that $y_4$ 
is positive for $n>N_c$ and negative in the opposite case, 
with $N_c$ a bit smaller than $3$ \cite{rev-01}.
The series (\ref{y4eq}) is alternating in signs for $n\alt 5$.
However, in this 
case, the coefficients are not so small for a simple Pad\'e summation to be 
effective.
Thus, we apply a PBL resummation.
Fig. \ref{fig} sketches the non-defective PBL approximants for $n=2,3$.
It is evident that several approximants are very close to each other, 
whereas the $[1/M]$ ones are well separated.
Usually, when facing with a similar situation, the PBL approximants that 
are far from the mean-value are discarded in the averaging procedure.
Since in this case we may not take advantage of the Pad\'e data,
in order to be sure  to not underestimate the error, we report in 
Table \ref{risultati}  two different estimates: 
one is the average over all PBL (which we will term ``safe''), 
the other (``best'') is obtained discarding the $[1/M]$ approximants. 
For $n=5,16$ we report only the ``safe'' estimate, because we do not have anymore evidence for a trend in the $[1/M]$ approximants.

\section{Conclusions}
\label{concl}

In this paper we determined the critical exponents associated with harmonic
operators of degree 2, 3, and 4 in O($n$) models by means of pseudo-$\e$
expansion. 
All our results for $y_i$, $\phi_i$, $\gamma_i$, and $\beta_i$ are 
reported for $n=$ 0, 1, 2, 3, 4, 5, and 16 in Table \ref{risultati}.

\begin{table*}[t]
\begin{center}
\caption{\small Final estimates from six-loop pseudo-$\e$ expansion. 
The values of $y_i$ are calculated directly from the series reported in 
the text. 
Whereas $\phi_i$, $\gamma_i$, and $\beta_i$ are obtained by 
means of scaling laws, using the most accurate theoretical estimates 
for standard critical exponents~\protect\cite{best} with our Pad\'e final results of the RG dimension $y_i$.}
\squeezetable
\begin{tabular}{l|ccccccc}
&$n=0$&$n=1$&$n=2$&$n=3$&$n=4$&$n=5$&$n=16$\\
\hline
 $y_2$ (PBL) &-&1.7348(2)& 1.7645(3)& 1.7897(3) & 1.8112(6) &1.830(1)&-\\
 $y_2$ (Pad\'e)    &-& 1.733(6)& 1.763(4)& 1.789(3) & 1.811(2) & 1.830(2)& 1.927(4)\\
$\phi_2$  &1& 1.092(4)  & 1.184(3)  & 1.272(2)  & 1.356(4)  & 1.398(4) &1.755(4) \\
$\beta_2$ &-&  0.798(4)  & 0.831(3)   &0.861(2)  & 0.891(3)    &  0.894(3) &0.978(4) \\
$\gamma_2$&-&0.294(8)& 0.353(5)  & 0.411(4)    & 0.466(3) & 0.504(3)  & 0.778(7)   \\
\hline
$y_3$ (PBL) &0.7258(2) & 0.8153(3) &0.8920(2) &0.956(2) &1.011(2)&1.059(4) &- \\
$y_3$ (Pad\'e)    & 0.725(30)  &  0.814(10)     & 0.891(9)   & 0.957(4)   &1.014(4)&
 1.063(7)&  1.31(1)\\
$\phi_3$  &0.426(18)   &  0.513(6)     &0.598(6)   & 0.681(3)  & 0.759(4)  & 0.812(6)  &1.19(1)\\
$\beta_3$ &1.336(18)  & 1.377(6)      & 1.416(6)    & 1.453(3)  & 1.488(5)  &
1.480(7)  &1.54(1)   \\
$\gamma_3$&  -0.91(4)  &-0.865(13) &-0.818(12)    &  -0.772(6)  & -0.728(6)  & -0.668(11)  &-0.35(2)\\
\hline
$y_4$ safe &-0.380(18)& -0.23(1)&-0.098(6)& 0.012(6)&0.104(8)&0.188(8)&0.62(1)\\
$y_4$ best &-0.393(5)&-0.236(3)&-0.103(1)&0.0094(30)&0.107(5)&&\\
\end{tabular}
\label{risultati}
\end{center}
\end{table*}

In order to make a comparison with the values in the literature
we report in Table \ref{compare} all the most accurate theoretical estimates 
for $y_i$, as obtained by means of scaling relations (\ref{scalrel}) 
using the most precise determinations of standard critical 
exponents \cite{best}.
For all values of $n$ our estimates are in perfect agreement with all 
known results and in the majority of the cases they are the 
most precise ones. We stress that such high accurateness should not be due to 
underestimation of the uncertainty, since we check our final error bars  
with other resummation techniques, such as Pad\'e-Borel-Leroy and 
conformal mapping. 
However the large $n$ results need to be discussed, since 
our estimates for $n\agt6$ can be affected by systematic errors, because 
the perturbative series we summed have not alternating signs. 
Anyway, for $n=16$ all the theoretical estimates are in good agreement, 
signaling that the evaluation of our uncertainty is probably good even in 
this case.

\begin{table*}[t]
\begin{center}
\caption{\small Theoretical estimates of the RG dimension $y_i$ as obtained 
by various approaches for several $n$:
five-loop $\epsilon$ expansion ($\e$ exp), 
six-loop fixed-dimension expansion (FD),
high-temperature expansion (HT exp), Monte Carlo simulations (MC),
and $1/n$ expansion at order $O(1/n^2)$ or $O(1/n)$.
The results are obtained by using scaling relations, using the 
most precise theoretical estimates for standard critical 
exponents \protect\cite{best}.}
\squeezetable
\begin{tabular}{l|ccccccc}
$y_2$&&$n=1$&$n=2$&$n=3$&$n=4$&$n=5$&$n=16$\\
\hline
6-loop (pseudo-$\e$)&&1.733(6)&1.763(4)  & 1.789(3) &1.811(2)  &1.830(2)&1.927(4) \\
6-loop (FD) \cite{cpv-02a}&&&1.763(18)& 1.787(30)&1.80(5) & 1.83(5)&1.92(6)\\
5-loop ($\e$-exp) \cite{cpv-02}&&&1.766(6)&1.790(3)&1.813(6)&1.832(8)\\
MC \cite{bfms-98} &&&1.755(3)& 1.787(3)& 1.812(2)\\
MC  \cite{Hu-01} &&& &&&1.815(39)\\
HT exp. \cite{PJF-74}&&& 1.750(22)& 1.758(21)&&\\
O($1/n$) \cite{note}&&&&1.640&1.730&1.784&1.932\\
O($1/n^2$)  \cite{G-02,note2}&&&&1.78(14)&1.83(6)&1.88(2)&1.95(3)\\
\hline
$y_3$&$n=0$&$n=1$&$n=2$&$n=3$&$n=4$&$n=5$&$n=16$\\
\hline
6-loop (pseudo-$\e$)&0.725(30)  &0.814(10)& 0.891(9)   &0.957(4)   &1.014(4)&1.063(7)&1.31(1)\\
6-loop (FD)\cite{dpv-03}&0.758(19)&&0.895(15)&0.953(23)&1.015(31)&1.065(19)&1.310(13)\\
5-loop ($\e$-exp) \cite{dpv-03}&0.739(9)&&0.892(22)&0.958(42)&1.020(45)&1.064(25)&1.28(10)\\
O($1/n$) \cite{note}&&&&&0.791&0.933&1.323\\
\hline
$y_4$&$n=0$&$n=1$&$n=2$&$n=3$&$n=4$&$n=5$&$n=16$\\
\hline
6-loop (pseudo-$\e$) best &-0.393(5)&-0.236(3)&-0.103(1)&0.0094(30)&0.107(5)&&\\
6-loop (pseudo-$\e$) safe &-0.380(18)& -0.23(1)&-0.098(6)& 0.012(6)&0.104(8)&0.188(8)&0.62(1)\\
6-loop (FD)  \cite{cpv-00,comm}
&&&-0.103(8)&0.013(6)&0.111(4)&0.189(10)&\\
5-loop ($\e$-exp)  \cite{cpv-00,cpv-02}
&&&-0.114(4)&0.003(4)&0.105(6)&0.198(11)&\\
MC Ref. \cite{ch-98}&&&-0.17(2)&-0.0007(29)&0.130(24)\\
O($1/n$) \cite{note}&&&&&&-0.08&0.662\\
\end{tabular}
\label{compare}
\end{center}
\end{table*}

Let us finally compare our values with some experiments.
We mention the result $\phi^{n=2}_2=1.17(2)$ for 
the ($2\rightarrow 1+1$) bicritical point in GdAlO$ _3$ \cite{rg-77}, and 
$\phi^{n=3}_2=1.279(31)$ in the  ($3\rightarrow 2+1$) study of 
MnF$ _2$ \cite{kr-79}.
Other experimental measures of $\phi_2$ can be found in Ref. \cite{priv}.
The experimental results obtained for a nematic-smectic-A transition 
reported in Ref. \cite{Wu-etal-94} are $\beta_2^{n=2}=0.76(4)$ and 
$\gamma_2^{n=2}=0.41(9)$.
For the third harmonic exponent we quote 
$\beta_3^{n=2} \simeq1.66$ in liquid crystals \cite{Aharony-etal-95},
$\beta_3^{n=2}=1.50(4)$ (Ref. \cite{nipt3}) 
$\beta_3^{n=2}=1.80(5)$ (Ref. \cite{nipt1}) in Rb$ _2$ZnCl$ _4$.
All these values compare well (within their own uncertainties) 
with our results.

\section*{Acknowledgment}

PC acknowledges financial support from EPSRC Grant No. GR/R83712/01.


\begin{references}
\bibitem{note2}
In Ref. \cite{G-02} were reported the four different estimates 
for $\phi_2$, $\beta_2$, $\gamma_2$, and $\beta_2+\gamma_2$ as obtained by
Pad\'e-Borel analysis of the corresponding $O(1/n^2)$ expansions.
These four estimates obviously leads to different estimates of $y_2$.
The value in the Tab. \ref{compare} is the average of the four $y_2$ as 
obtained by means of scaling relations (for the exponent $\nu$ we take
the value in \cite{best}) and the uncertainty is the maximum deviation.
Obviously, the quoted error bars are  only indicative.
\end{references}
\end{document}